# Mean Switching Frequency Locking in Stochastic Bistable Systems Driven by Periodical Force

Boris Shulgin,, Alexander Neiman, Vadim Anishchenko

*Department of Physics, Saratov State University, Astrakhanskaya St. 83, 410071, Saratov, Russia*

(October 17, 1995)

The nonlinear response of noisy bistable systems driven by strong amplitude periodical force is investigated by physical experiment. The new phenomenon of locking of the mean switching frequency between states of bistable system is found. It is shown that there is an interval of noise intensities in which the mean switching frequency remains constant and coincides with the frequency of external periodic force. The region on the parameter plane "noise intensity − amplitude of periodic excitation" which corresponds to this phenomenon is similar to the synchronization (phase–locking) region (Arnold's tongue) in classical oscillatory systems.

The dynamics of noisy nonlinear systems show a variety of non-trivial phenomena which have been extensively studied during the last decade [1,2]. Among them the resonance-like and synchronization-like phenomena are of great interest [3]. In particular, a great deal of work has been devoted to the phenomenon of stochastic resonance (SR) [4]. This phenomenon occurs in nonlinear systems subjected simultaneously by external noise and periodic force. The response of nonlinear noisy system to small periodic excitation can be enhanced. It happens most effectively for an optimal intensity of stochastic force when the noise-controlled time scale (for example, the mean transition time between two stable states of a bistable system) coincides with the time scale of the periodic force. Theoretical investigations (see references in [5]) have shown that SR phenomenon can be correctly described in terms of linear response theory (LRT) [6].

The stochastic synchronization has been observed in two coupled bistable systems [7]. It was found out that when the strength of coupling achieves some critical value then the stochastic hopping dynamics in the subsystems becomes coherent. In papers [8] the resonance phenomena in globally coupled stochastic oscillators have been studied.

Nonlinear effects in stochastic resonance have been studied in [9–14]. In [12] an analytic approach in the framework of the adiabatic theory [15] has been proposed. The generation of high-order harmonics has been considered in [11]. In [9] the universal power low decay of the spectral density has been found in the weak–noise limit. Another nonlinear effect in SR, noise–enhanced heterodying has been described in [10]. A new nonlinear effect in SR has been found and accounted theoretically in [14]: using the technique of pulse sequences the existence of a second peak in the dependence of the signal–to–noise ratio versus the noise intensity has been found for a large enough amplitudes of periodic force.

In the present Letter we study another group of nonlinear phenomena in periodically driven noisy bistable systems. Let us turn to the classical theory of oscillation. As is well known, small periodic forcing of an oscillator leads to the phenomenon of linear resonance: when the driven frequency coincides with the natural frequency of oscillator then the magnitude of response of the system takes its maximum. The phenomenon of SR is similar to this linear resonance. As distinct from ordinary resonance the natural frequency in the case of SR is the statistical quantity and the phenomenon is observed via changing of this noise-controlled quantity.

Another resonance phenomena are observed in self–sustained oscillators: the natural frequency of oscillator can be locked by external periodic force. As a result, regions of synchronization in the parameter space of the systems appear. These regions are called "Arnold's tongues", and the natural frequency of oscillator is in a rational relation with the driving frequency in these regions. It is reasonable to try to find similar phenomena in periodically driven stochastic bistable systems. Actually, a bistable system driven by external noise can be considered as an analog of self–sustained oscillator with natural frequency represented by the mean switching frequency (MSF) between stable states. We set up the hypothesis that in nonlinear regime of operation of stochastic bistable system driven by periodic force the same regions where the MSF coincides with the frequency of driving force can be observed. Below we show that this hypothesis fits experimental data.

As was mentioned above, the effects under consideration are sufficiently nonlinear and therefore the existent theories of periodically driven stochastic systems can not be applied. We choose physical experiment as a technique for investigations. The two models we used are the Schmitt trigger and an overdumped bistable oscillator.

The Schmitt trigger is an ideal two–state electronic device demonstrating pure hopping dynamics. Using this device the stochastic resonance was firstly investigated experimentally in [16]. A schematic diagram of the Schmitt trigger system and description of its operation can be found for instance in [16,15,5]. The application of the adiabatic theory to this device has been made in [15]. The ideal Schmitt trigger circuit driven by periodic force and noise $\xi(t)$ obeys the equation

$$y = sgn[\gamma y - A\cos(2\pi f_0 t) - \xi(t)], \qquad (1)$$

where $\gamma$ is the parameter corresponding to the threshold levels of the trigger.

The overdumped bistable oscillator simultaneously driven by noise and periodic signal is described by the Langevin equation

$$\dot{x} = ax - bx^3 + A\cos(2\pi f_0 t) + \xi(t), \quad (2)$$

where $a$ and $b$ are parameters. In the absence of noise bistability is destroyed for $A \geq V_b = (4a^3/27b)^{1/2}$. The parameter $V_b$ is equivalent to the threshold level of the Schmitt trigger. The detailed description of experimental investigation of this system has been done in [17].

In our experiments we use quite the same schemes as in cited papers. The Schmitt trigger which is just an operational amplifier is subjected by noisy signal with cut-off frequency $f_c = 100kHz$ and the periodic signal. The amplitude of the periodic signal $A$ in all experiments is small enough not inducing switching of the trigger without noise: $A < V_t$, where $V_t = 150mV$ is the threshold of the Schmitt trigger. At the output of the Schmitt trigger system we have got the dihotomic stochastic process which can be characterized by the mean durations of the upper state and lower state: $T_u$, $T_l$. We calculate these quantities using a computer connected via ADC with the output of the system. The mean "period" of switching is therefore

$$T_s = T_u + T_l. \quad (3)$$

In the frequency domain this quantity corresponds to the mean switching frequency (MSF)

$$f_s = \frac{1}{T_s} = \frac{1}{T_u + T_l} \quad (4)$$

In the absence of periodic force the MSF is fully controlled by noise and is characterized by the exponential Arhenius law [2]:

$$f_s^{(0)} \propto \exp(-\Delta U/D), \quad (5)$$

where $\Delta U$ is a barrier height and $D$ is the noise intensity. In the presence of periodic excitation the MSF becomes a function of the parameters of periodic force.

The results of measurements of the MSF for the Schmitt trigger are shown on Fig.1 as a function of noise intensity. In the absence of periodic excitation as well as for a weak periodic forcing the dependence of the MSF versus noise intensity fits exponential law. For a large enough amplitude of periodic force the exponential law breaks down. It is seen that there is an interval of noise intensities in which the MSF remains constant and corresponds to the frequency of periodic force $f_0$. The variations of the MSF in this region do not exceed $\pm 0.5\%$. Therefore, the mean switching rate between two states of noisy bistable system is "locked" by external periodic force: in a certain region the MSF is equal to the value of driving frequency.

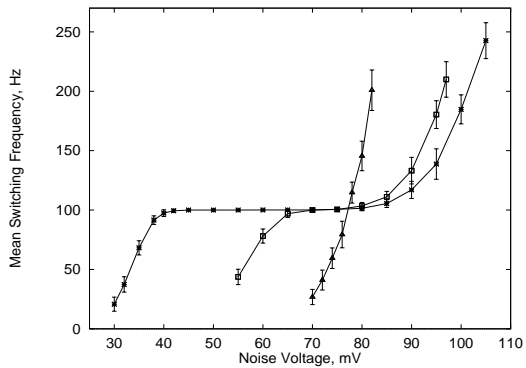

FIG. 1. The measured MSF versus noise voltage for different amplitudes of periodic signal for the Schmitt trigger: $A = 0mV$ ($\triangle$), $A = 60mV$ ($\square$), $A = 100mV$ ($\star$). The signal frequency is $f_0 = 100Hz$ and the trigger threshold level is $V_t = 150mV$.

Making the similar measurements for different values of the amplitude of periodic force we obtain the region on the parameter plane "noise intensity – amplitude of periodic force" in which the MSF is equal to the frequency of periodic force within the limits of experimental accuracy given above. These "synchronization" regions are shown on Fig.2 for several values of driving frequency $f_0$. The base of each of the regions determines the synchronization threshold values $A_{th}$ of modulation amplitude. Therefore, the phenomenon has threshold feature as in classical oscillators with hard excitation. Fig.2 demonstrates the dependence of the synchronization threshold values $A_{th}$ versus driving frequency and noise intensity as well: the greater the driving frequency is, the greater the threshold value $A_{th}$ is and the stronger noise we have to apply to obtain the effect of the MSF locking.

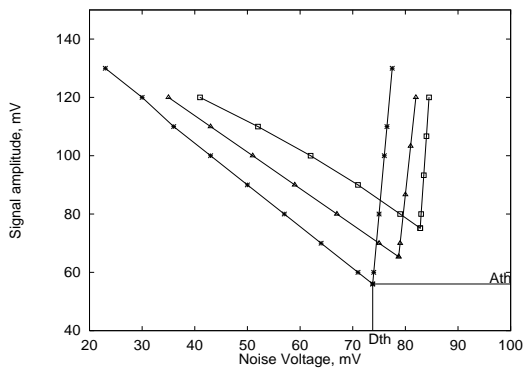

FIG. 2. The synchronization regions for the Schmitt trigger for different frequencies of periodic force: $f_0 = 100Hz$ ($\star$), $f_0 = 250Hz$ ($\triangle$), $f_0 = 500Hz$ ($\square$). The threshold level of the trigger is $V_t = 150mV$.

The same regions of synchronization as in Fig.2 can be obtained via varying the threshold levels of the Schmitt trigger which are presented on Fig.3. This figure shows the dependence of the synchronization threshold value on the threshold levels of the Schmitt trigger: the higher the barrier height of the trigger is, the greater the threshold value $A_{th}$ and the noise intensity $D_{th}$ are.

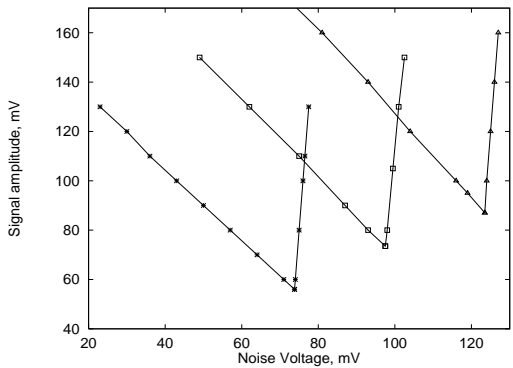

FIG. 3. The synchronization regions for the Schmitt trigger for different values of the trigger threshold level: $V_t = 150 mV$ ($\star$), $V_t = 205 mV$ ($\square$), $V_t = 255 mV$ ($\triangle$). The frequency of periodic force is $f_0 = 100 Hz$.

The measurements of the signal–to–noise ratio (SNR) inside the regions of synchronization have shown the existence of an additional maximum in the dependence of the SNR versus the noise intensity. This effect is in full correspondence with the results of the paper [14]. Below the threshold value $A_{th}$ (see Fig.2), where the amplitudes of the periodic force are small enough, the dependence of the SNR versus the noise intensity has ordinary shape with a single maximum.

Qualitatively the same phenomena have been observed for the overdamped bistable oscillator. The results of measurements of the MSF for different amplitudes of periodic force are presented on Fig.4. Again we observe a region of the noise intensity in which the MSF remains constant. The results of experiments were confirmed by numerical simulations of equations (1) and (2).

FIG. 4. The measured MSF versus noise voltage for the overdamped bistable oscillator for different amplitudes of periodic signal: $A = 0 mV$ ($\triangle$), $A = 480 mV$ ($\square$), $A = 620 mV$ ($\star$). The signal frequency is $f_0 = 100 Hz$. Bistability is destroyed for the threshold parameter $V_b = 630 mV$.

The "synchronization" regions on Fig.2,3 are very similar to those in a classical self-sustained oscillator driven by external periodic force (Arnold's tongues). However there is a basic difference between phase–locking effects in self–sustained oscillators and the phenomenon of the MSF locking. In our case of stochastic bistable system there is no any natural frequency in classical sense. The role of natural frequency of oscillator is played by the statistical quantity, the mean switching frequency between two states of the system. The notion of a "phase" for this stochastic switching is difficult to introduce. Consequently, we can mark the phenomenon under consideration as a synchronization only in quotation marks.

**In conclusion**, we have studied experimentally the nonlinear effects in bistable stochastic systems driven by periodic force. We have found out the new phenomenon: the mean switching frequency locking by external periodic force. This phenomenon manifests itself in the existence of a region on the parameter plane "noise intensity – amplitude of periodic force" in which the mean switching frequency between two states of the system equals to the frequency of periodic force. This synchronization-like phenomenon has threshold features. The threshold value of the signal amplitude depends on the signal frequency as well as on the barrier height of the potential.

The authors would like to thank P. Hänggi, A.R. Bulsara, L. Schimansky-Geier, E. Pollak, D. Postnov, J. Kurths, T. Vadivasova and V. Astakhov for valuable discussions. This work was supported, in part, by the International Scientific Foundation grants RNO000 and RNO300.

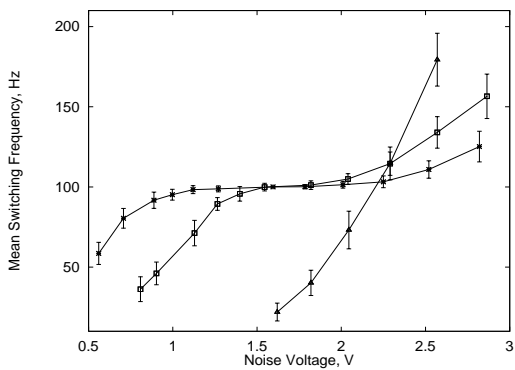

4